\documentstyle[amssymb]{article}

\newcommand{\tensor}{\otimes}
\newcommand{\defeq}{\stackrel{\mbox{\tiny {def}}}{=}}
\newcommand{\real}{{\Bbb R}}

\newcommand{\compl}{\Bbb C}
\newcommand{\comm}[1]{}
 
\newtheorem{theorem}{Theorem}
\newtheorem{lemma}[theorem]{Lemma}
\newtheorem{corollary}[theorem]{Corollary}

\title{Multicomponent WKB on arbitrary symplectic manifolds: 
A star product approach}
\author{C. Emmrich, H. R\"omer\\
Fakult\"at f\"ur Physik
\\Universit\"at Freiburg   \\
Hermann-Herder-Stra\ss e 3\\
79104 Freiburg}
\date{}

\begin{document}
 
\maketitle

\begin{abstract}
It is known that in the WKB approximation of multicomponent
 systems like Dirac equation or Born-Oppenheimer 
approximation, an additional phase appears apart from the
 Berry phase. So far, this phase was only 
examined in special cases, or under certain restrictive 
assumptions, namely that the eigenspaces of the matrix or 
endomorphism  valued symbol of the Hamiltonian form trivial
 bundles.

We give a completely global derivation of this phase which
 does not depend on any choice of local trivializing 
sections. This is achieved using a star 
product approach to quantization. Furthermore, we give a
 systematic and global approach to a reduction of the 
problem to a problem defined completely on the different
 ``polarizations''. Finally, we discuss to what extent it
 is actually possible to reduce the problem to a really
 scalar one, and make some comments on obstructions to the 
existence of global quasiclassical states.    
\end{abstract} 

\section{Introduction}
 
Whereas the semiclassical and WKB approximation of systems
 with Hamiltonians whose symbol is a scalar function is 
very well understood both on a local and a global level,
  this is not the case to the same extent for systems with
 Hamiltonians whose symbol are matrix valued functions on
 a symplectic manifold, or sections 
in the endomorphism bundle of a vector bundle over a
 symplectic manifold $M$. 

Such Hamiltonians appear in many places in physics: 
 Examples are the  Dirac equation, multicomponent wave 
equations like electrodynamics in media and Yang-Mills
 theories, and the Born-Oppenheimer approximation
in molecular physics.

In the scalar case,   a nice geometric interpretation of the 
semiclassical states as half densities on Lagrangian 
submanifolds invariant under the Hamiltonian flow exists,
 and discrete spectra may be computed 
using the Bohr-Sommerfeld condition, taking into account 
the Maslov correction.

In the multicomponent case, the analogous structures are 
not known:
There exist many results on a local level, but on a global 
level results are only known either for special examples, 
or under the very restrictive assumption that the 
eigenspaces  of the symbol of the Hamiltonian form trivial 
bundles over the symplectic base manifold $M$ 
\cite{ma-fe:semiclassical,klein,li-fl:geometric}.

Locally, the WKB ansatz for multicomponent systems still is 
 \[\Psi(x) = a(x,\hbar) \exp(i S(x)/\hbar)\]
 with  $a(x,\hbar) = a_0(x) + a_1(x)\hbar +...$. 
Here, the $a_i(x)$ are now vector valued functions
\cite{gu-st:geometric,ka-ma:operators,ma-fe:semiclassical}.

In lowest order in $\hbar$ the time independent 
Schr\"odinger equations yields:
\[\bigl[ H_0\bigl(x, d S(x)\bigr) -E \bigr] a_0(x) = 0,\]
i.e.,  $a_0(x)$ has to be an eigenvector of the matrix
 valued principal symbol $H_0(x,p)$ at the point
 $p = dS(x)$. Denoting by $\lambda_\alpha(x,p)$ 
the $\alpha$-th eigenvalue of the matrix $H_0(x,p)$ at 
the point  $(x,p)$, the phase $S(x)$ has to satisfy the
Hamilton-Jacobi-equation for one of the eigenvalue functions 
$\lambda_\alpha$:  $~~ \lambda_\alpha(x, dS) -E  =0$,
and one obtains a separate phase function for each 
``polarization'' $\alpha$.

The equation of order $\hbar$ has a more complicated
 structure than in the scalar case: The absolute value of
 the amplitude still fulfills the usual transport equation,
 but the vector $a_0$ satisfies a more complicated equation
 which contains a Berry phase term, and an additional term
 without obvious geometric meaning
\cite{yabana,li-fl:geometric,ka-ma:operators}. 

There exist some functional analytically rigorous results on
the WKB approximation for special  systems in the literature
\cite{ma-fe:semiclassical,klein}. However, no analog of the 
Bohr-Sommerfeld quantization condition is known in general,
 and all those results have the common feature that they 
are based on the assumption that the bundles of eigenvectors
 of the symbol of the Hamiltonian are trivial, at least on 
the ``classically allowed region'' 
$\lambda_\alpha< \tilde{E}$ for an $\tilde{E}>E$.
This assumption is not fulfilled in many 
physical applications: Apart from cases like the Dirac 
equation on topologically non-trivial space-times, 
where the possibility 
of non-trivial eigenbundles is quite obvious, 
this phenomenon even arises in applications to the 
Born-Oppenheimer approximation:
Although the nuclear phase space is contractible, 
non-trivial bundles may arise due to the phenomenon of level
 crossing, i.e., the possibility of two  eigenvalues 
becoming equal in certain points on phase space: 
The degeneracy increases in those points. For the WKB 
approximation one has to cut out those points and to treat
 them separately. This may lead to the appearance of 
non-trivial bundles, and as a matter of fact, it
is well known that non-trivial bundles play a role in 
real molecular systems.

In \cite{li-fl:geometric} a Bohr-Sommerfeld condition 
could be derived  for systems whose symbol has 
non-degenerate eigenvalues, using 
the Moyal  product on matrix valued symbols in order 
to reduce this case to the WKB approximation for a 
scalar system. 
To this end, a formally unitary $U=U_0 + U_1 \hbar +....$ 
was introduced such that 
\[ U^\dagger * H * U = \Lambda \]
for some diagonal matrix valued symbol $\Lambda$.

However, this method has several disadvantages: First,
considering this equation at zeroth order, one concludes 
that the columns of $U_0$ are eigenvectors of $H_0$. Hence, 
this method may   again be applied globally only if the 
eigenvector bundles are trivial. Moreover, it is not 
possible to generalize this method to the 
case of degenerate eigenvalues with constant degeneracies. 
In this case, it is in general  not possible to diagonalize 
the Hamiltonian globally (even if the eigenvector bundles 
are trivial), but only 
to block-diagonalize it. Finally, this approach is not 
natural from a geometric point of view: The Moyal product is 
defined only on trivial bundles over $\real^{2n}$, hence it 
is not possible to stick with the Moyal
product if one wants to reduce the problem in a global 
way to a problem defined on the respective eigenbundles.
 In particular, it will become obvious that this method, 
if generalized to the degenerate
case, will correspond to a quantization based on a 
non-canonical and unnatural choice of a connection: 
The choice of $U_0$ implies, as mentioned above, a choice 
of trivializing sections of the eigenbundles.  
In a sense to become clear in section \ref{sec:corrections}, 
the quantization achieved by this process corresponds to 
a connection with respect to which the chosen set of 
trivializing sections  is covariantly constant. 
Such a connection is obviously geometrically
meaningless,
 since it depends on the choice of those sections. 

The goal of this article is  to reduce the problem of WKB 
for endomorphism valued symbols ``as far as possible'' 
to the scalar case, using only globally defined methods:
 Apart from being an important progress in a geometric 
understanding of multicomponent WKB approximation,  
this is an important step towards a Bohr-Sommerfeld like 
condition, since the latter is necessarily global in nature. 
 
For this reduction of the problem, the notion of star 
products on arbitrary endomorphism bundles proves to be 
particularly useful, since we have to deal with  non-trivial
 bundles. Such star products are
known to exist on arbitrary endomorphism bundles over a 
symplectic manifold, as has been
shown by Fedosov \cite{fedosov}. His method is particularly 
adapted to our purposes,  since his  proof of
existence is constructive in nature, and there exist 
explicit formulas for the isomorphisms of different star 
products.
Although we will restrict ourselves to formal star products, 
i.e., to formal power series in $\hbar$,
 the method can in principle be translated into operator 
language using the notion of asymptotic operator 
representations,
 which are, in a suitable sense, representations up to order 
$O(\hbar^\infty)$ of the star product algebra.
Such representations are known to exist if certain 
quantization  condition are satisfied \cite{fedosov2}.

The main idea of the paper is to give a 
``basis independent and geometrically natural version 
of the diagonalization procedure of \cite{li-fl:geometric}''.
 What is meant by this becomes more apparent if one 
observes that essentially all additional complications 
 in the WKB approximation of multicomponent systems are 
due to the fact that the projection operators onto the 
eigenspaces of the principal symbol of $H$ are in general 
not constant. If one could replace ordinary derivatives by
 covariant derivatives with respect to which the projections
 are covariantly constant, the problem 
would be considerably simplified. However, as we assume 
that we are studying  a given definite quantum mechanical
 system, we cannot simply replace ordinary derivatives by 
covariant derivatives ``by hand'', as we do not want to 
change the physics of the system. 

Here, the use of Fedosov's star product proves to be 
particularly powerful: His construction
of a star product depends on the choice of a symplectic 
connection on the base space
and a connection on the endomorphism bundle. Nevertheless, 
the different star products 
are isomorphic. Hence, if we interpret the choice of the 
connection as defining a definite
quantization procedure, then we may change this connection 
without changing physics if we apply at the same time the 
corresponding isomorphism. This means
that we are allowed to change the connection in order to 
simplify the problem, if we consistently correct the 
symbols of the Hamiltonian and  all interesting observables 
at the same time. 

Indeed, this additional freedom (which is completely 
missing if one restricts oneself 
to the Moyal product) may be used to simplify the problem 
considerably: One may find 
an $\hbar$-dependent connection, $\hbar$-dependent 
projections, and a corrected symbol
such that all these structures are compatible in an 
obvious sense.
 An important feature
of Fedosov's product is that the star product constructed 
from these data is compatible 
with the projections as well, i.e,. if $A$ and $B$ are 
observables such that there 
corrected symbols commute with the projections, then the 
corresponding star product  
$A*B$ commutes with the projections as well. Hence, the 
different blocks defined 
by the projections are decoupled for a suitable set of 
observables.

Using these techniques, we  will be lead to a geometrical
 interpretation of the additional phases appearing in the 
WKB approximation  at order 
$\hbar$ as formed from the Poisson-curvature and an analog 
of the second fundamental form of a Berry-type connection. 
We will show that this even
holds if one chooses a different connection for computing 
the correction. 
Hence, the Berry type connection is distinguished among the 
set of all compatible connections.

The paper is organized as follows: In the next section we 
give a short overview over Fedosov's
construction of star products on arbitrary endomorphism 
bundles to the extent needed for our
applications. The main fact will be that a modification of 
the connection  on the endomorphism
bundle at some order $O(\hbar^k)$ induces an isomorphism of 
the corresponding star product 
algebras which implies corrections only at higher order 
$O(\hbar^{k+1})$. This will allow the use
of inductive methods in the main part of the paper, and 
to compute the corrections iteratively
order by order in $\hbar$.

In section \ref{sec:block_diag}, we show how one may use 
the freedom of choosing the connection
in order to block-diagonalize the system, i.e., to find a 
decomposition of the vector bundle
such that the Hamiltonian and a large class of observables 
preserve this decomposition at
the quantum mechanical level. Furthermore, we will show 
that for a suitable class of 
observables, a  formal Heisenberg equation of motion may 
be defined.

In section \ref{sec:corrections} we compute the correction 
of the symbol of $H$ at order
$\hbar$ for all compatible connections, thus reproducing in
 particular the additional phase observed in special cases 
before. 
 We give a geometric interpretation of these terms, 
and show that a certain adapted  connection which is of 
``Berry type'' is distinguished among all adapted
 connections. 

Finally, in section \ref{sec:scalar} we examine to which 
extent it is actually possible to reduce the problem to a
 scalar one, and discuss some 
aspects connected to a Bohr-Sommerfeld like quantization
 condition.

\section{Star products on endomorphism bundles}

In this section we give a short introduction to Fedosovs
 approach to deformation quantization \cite{fedosov}, as
 needed for our purposes,
and we prove three lemmas needed in section
 \ref{sec:block_diag}

We assume the reader to be familiar with the general ideas
 of deformation quantization. For an introduction, we refer
 to \cite{bayen}. 

We denote the Moyal product, which is the star product on
 $\real^{2n}$ corresponding to Weyl ordering, by $\diamond$: 
\[ (a\diamond b)(x) = ( e^{\frac{i\hbar}{2} \omega^{kl}
\frac{\partial}{\partial x^k} \frac{\partial}{\partial y^l}}
a(x) b(y) ) \rule[-2mm]{0.1mm}{6mm}_{x=y}, \]
where $\omega^{kl} $ are the components of the Poisson
 tensor. 
The algebra ${\cal F}(\real^{2n}) \otimes \compl[[\hbar]]$
 with the product $\diamond$ is called the 
{\em Weyl algebra} over $\real^{2n}$.
(Here, ${\cal F}(\real^{2n})$ denotes the algebra of smooth
 functions on $\real^{2n}$.)

Let $M$ be a symplectic manifold with symplectic form
 $\omega$ and a symplectic connection $\partial_s$ 
(such a connection always exists), and let
$V$ be a hermitian vector bundle over $M$ with hermitian
 connection $\nabla$. 
We consider the bundle $V[[\hbar]] = V \tensor
 \compl[[\hbar]]$
which is a fiber bundle over $M$ whose fiber over a point
 $q \in M$ is the $\compl[[\hbar]]$ module 
$V_q[[\hbar]] =V_q \tensor \compl[[\hbar]]$.
 Let $E[[\hbar]]= \bigcup_{q \in M}End(V_q)
 \tensor \compl[[\hbar]]$
   be the bundle of endomorphism s of  the fibers of
 $V[[\hbar]]$.  

We may define a star product on the  endomorphims  
 $\Gamma(E[[\hbar]])$ of the  bundle $V[[\hbar]]$ by 
Fedosov's construction \cite{fedosov}. To this end one 
considers the algebra 
\[{\cal A} = \Gamma( E[[\hbar]] \tensor W),\]
 where $W$ is the Weyl algebra bundle over $M$ 
 (i.e., the bundle over $M$ whose fiber over $q$ consists 
of Weyl algebra on  the symplectic vector space $T_q M$). 
 
${\cal A}$  is graded as a vector space, with a degree
 defined by:
\[ deg(\hbar)=2, ~~~ deg(y^i) =1 , \]
where the $(y^i)$ are coordinates on the fiber $T_q M$.
Defining ${\cal A}_i$ as the set of all $\hat{X} \in {\cal A}$
 which are sums of terms of degree larger or equal $i$,
 ${\cal A}$ becomes a filtered algebra:
\[ {\cal A}={\cal A}_0 \supset {\cal A}_1 \supset \ldots .\]

The hermitian connection $\nabla$ on $V$ induces a 
connection on $E[[\hbar]]$ which we denote by $\nabla$ 
as well. Together with the  symplectic
connection $\partial_s$ on $M$ this  defines a connection
 $\partial =
1 \tensor \partial_s  + \nabla \tensor 1$ on $E[[\hbar]] 
\tensor W $.

One may construct by an iterative procedure, starting with
 $\partial $, an 
exterior covariant derivative $D$ on ${\cal A} \tensor
 \Omega(M)$, the forms on $M$ with values in 
${\cal A}$, such  that

\begin{enumerate} 
 \item      $D^2 =0 $
 \item \label{it:isom}  
the space of covariant constant elements in ${\cal A}\hat{=} 
{\cal A}\otimes \Omega_0(M) \subset{\cal A}\otimes \Omega(M)
 $ is isomorphic as a  vector space to $E[[\hbar]]$ 
\label{en:isom}  
\item $D$ acts as a derivation on 
${\cal A} \tensor  \Omega(M)$  
(with the product  induced by the Moyal product on 
${\cal A}$).  
\end{enumerate}

These properties guarantee that the covariant constant 
sections form a subalgebra and  allow us to use the  vector
 space isomorphism of \ref{it:isom}. 
between the covariant constant elements in ${\cal A}$ and 
$E[[\hbar]]$  
to define the star product on $\Gamma(E[[\hbar]])$ by 
pull back of the product on ${\cal A}$. More precisely, 
if we denote the isomorphism from $\Gamma(E[[\hbar]])$ to
 the covariantly constant elements 
in ${\cal A}$ by ${\cal Q}$, and its inverse by $\sigma$, 
the star product on $E[[\hbar]]$ is given by:
\[ a * b = \sigma( {\cal Q}(a) \diamond {\cal Q}(b)) \] 
for $a,b \in \Gamma(E[[\hbar]])$. 

To give an explicit expression for this flat derivative, 
we introduce operators $\delta$ and $\delta^{-1}$ on 
${\cal A} \tensor \Omega(M)$:

\[ \delta = d x^i \wedge \frac{\partial}{\partial y^i} \]
and
\[ \delta^{-1} a = \frac{1}{p+q} i_{y^i \frac{\partial}
{\partial x^i}} a \]
if $a$ is a q-form that is homogeneous of degree $p$ in 
the fiber coordinate
$y$, $p +q \not = 0$, and $\delta^{-1} a = 0 $ if $p = q =0$.
These operators have the important property that for any 
$\hat{X} \in \Omega(M)\tensor {\cal A}$:
\[ \hat{X} = \hat{X}_{00} + ( \delta^{-1} \delta +  
\delta \delta^{-1}) \hat{X} \]
Here, $\hat{X}_{00}$ denotes the part of $\hat{X}$ that is a 
zero form  of degree $0$ in $y$.

The ansatz 
\[ D = - \delta + \partial + \left[\frac{i}{\hbar} r,
 ~~ \right] \]
for a one form $r$ with values in ${\cal A}$, 
the flatness condition $D^2 =0$ together with a 
``normalization condition'' $\delta^{-1} r = 0$ leads to 
the condition 
\begin{equation}
 r = \delta^{-1} \tilde{R} + \delta^{-1} \left( \partial r +
\frac{i}{\hbar}  r^2 \right), \label{eq:r_def}
\end{equation}
by which $r$ is uniquely defined and may be computed 
iteratively.
Here, $\tilde{R}$ is the curvature of the connection $\partial$,
considered as a two form with values in ${\cal A}$, acting on
$\cal A$ as $\frac{i}{\hbar} [ \tilde{R}, \cdot] $: If
$R^E$ is the curvature of $\nabla$ and $ R^s$ is the curvature of
$\partial_s$, then $\tilde{R}$ is locally given by:
\[ \tilde{R} = - i  \frac{\hbar}{2} R^E_{ij} d x^i \wedge d x^j +
 1 \tensor \frac{1}{4} \omega_{ij}(R^s)^i_{klm} y^j y ^k d x^l 
\wedge dx^m. \]
We note that in spite of the factor $1/\hbar$ in $(\ref{eq:r_def})$
the solution of this equation is a power series in $\hbar$, 
containing no negative powers.

With $r$ determined, the covariantly constant continuation 
${\cal Q}(a)$
of $a$ in ${\cal A}$ is given by the unique solution to the equation
\begin{equation} \label{eq:cov_const}
 {\cal Q}(a) = a_0 + \delta^{-1}( \partial {\cal Q}(a) + 
[ \frac{i}{\hbar} r,  {\cal Q}(a)]),  
\end{equation}
which may again be solved iteratively.

Different choices of $\nabla $ will yield different star products. 
However, they are equivalent in the following sense:

\noindent
Given two hermitian connections  $\nabla^{(1)}, \nabla^{(2)}$
and two symplectic connections 
$\partial_s^{(1)}, \partial_s^{(2)}$ with corresponding derivates
$D^{(1)},D^{(2)}$, there is a formally unitary 
\[U_{(12)} = \exp_{\diamond } (\frac{i}{\hbar }A_{(12)})\]
  for some $A \in {\cal A}$ such that:
 \[ D^{(1)} \hat{X} = 0 \Rightarrow D^{(2)}( U_{(12)} \diamond 
 \hat{X} \diamond  U_{(12)} ^{-1})
 = 0 \] for any $\hat{X} \in \Gamma({\cal A)}$. 
(Here, $ \exp_{\diamond }$ denotes the exponential with 
respect to the fibrewise Moyal product $\diamond $.)

Furthermore, letting ${\cal Q}^{(i)} (X) $ denote  the covariant constant
 continuation of $X  \in 
\Gamma(E[[\hbar]])$ with respect to $D^{(i)}$, and  
$\sigma^{(i)}$ denote the inverse of ${\cal Q}^{(i)} 
 $ ($ i =1,2$), we have an 
isomorphism of the different star-products given by:
 \[ \phi(X) \defeq  \sigma^{(2)}( U_{(12)}\diamond  {\cal Q}^{(1)}
 (X) \diamond  U_{(12)}^{-1})\] 
To check this statement one simply has to compute: 
\begin{eqnarray*}   
      \phi ( X *_1 Y ) 
& = &\sigma^{(2)}(U_{(12)}
\diamond{\cal Q}^{(1)}(X *_1 Y) \diamond  U_{(12)}^{-1})\\ 
 &=&
  \sigma^{(2)}(U_{(12)}\diamond   {\cal Q}^{(1)} (X ) 
   \diamond {\cal Q}^{(1)} (Y ) \diamond  U_{(12)}^{-1})\\
 &=&
  \sigma^{(2)}(
(U_{(12)}\diamond   {\cal Q}^{(1)} (X )\diamond  U_{(12)}^{-1}) 
   \diamond (U_{(12)}\diamond{\cal Q}^{(1)} (Y )
 \diamond  U_{(12)}^{-1}))\\
 &=&   
   \sigma^{(2)}(  {\cal Q}^{(2)}(\phi(X) )  \diamond 
 {\cal Q}^{(2)}   (\phi(Y))  \\ &=&  \phi(X) *_2 \phi(Y)
\end{eqnarray*}

We note that, in general, $\phi(X)\neq X$. This is precisely what 
was meant in the introduction by the statement that we can change
the quantization procedure without changing physics,
 if we consistently correct the symbols of the Hamiltonian and the relevant 
observables at the same time.

We may express $ U_{(12)} \diamond  \hat{X} \diamond  
U_{(12)} ^{-1}$  directly in terms of $A_{(12)}$  as:
\begin{equation} \label{eq:U_conj}
U_{(12)}\diamond  \hat{X} \diamond  U_{(12)}^{-1} = 
\sum _{k=0}^\infty \left(\frac{i}{\hbar}\right)^k \frac{1}{k!}
\underbrace{[A_{(12)},[A_{(12)},\ldots[A_{(12)},}_{\mbox{k-times}}
 \hat{X}]\ldots]] 
\end{equation}

For $A_{(12)}$, one may derive the following equation which 
defines it iteratively \cite{fedosov}:
\begin{equation} \label{eq:A_def}
 A_{(12)} = \delta^{-1} \left( \frac{ ad (\frac{i}{\hbar} A_{(12)})} 
{\exp_\diamond (ad(\frac{i}{\hbar} A_{(12)}) )-1 }
 (\Delta \gamma)_{(12)} +
 (D + \delta ) A_{(12)}   \right) 
\end{equation}
Here, $(\Delta  \gamma)_{(12)}$ denotes  the one form on $M$ with
 values in $\cal A$ (acting by 
$\frac{i}{\hbar} [(\Delta  \gamma)_{(12)} ,\cdot]$) 
which is induced by the difference of the two connection
forms corresponding to 
$\partial^{(i)} =
1 \tensor \partial_s^{(i)}  + \nabla^{(i)} \tensor 1,
 ~ i=1,2$ on $E[[\hbar]] \tensor W $. (The difference of two 
connection forms is tensorial and hence defines a one form on $M$).
If $\nabla^{(i)}$ is locally given by $ \frac{\partial}{\partial
 x^i } + \Gamma_i$, and the Christoffel symbols corresponding to
$\partial_s^{(i)}$ are  given by $(\Gamma_{s}^{(i)})^k_{lm}$,
then $(\Delta  \gamma)_{(12)}$ is locally given by
\begin{equation} \label{eq:Del_gam}
 (\Delta  \gamma)_{12} = -i \hbar  (\Gamma_i^{(2)} 
   -\Gamma_i^{(1)}) d x^i 
 +1\otimes \frac{1}{2} \omega_{ij}\left((\Gamma_{s}^{(2)})^i_{kl}
- (\Gamma_{s}^{(1)})^i_{kl}\right) y^j y ^k d x^l  
\end{equation}
In particular, it is obvious from this expression that $A_{(12)}$ 
does not contain any negative powers of $\hbar$.

\medskip

\noindent {\em Remarks:}

1.) Though $U$  does contain inverse powers of $\hbar$, 
 $ U_{(12)}\diamond  \hat{X} \diamond  U_{(12)}^{-1}$ is again
 a power series in $\hbar$
for all $\hat{X}  \in \Gamma({\cal A)}$ with $D^{(1)} \hat{X}=0$.
 This is shown in \cite{fedosov} by showing first that 
$ D^{(2)} (U_{(12)}\diamond  \hat{X} \diamond  U_{(12)}^{-1})=0$
in an enlarged algebra $W^+$ whose elements may  contain negative
 powers of
$\hbar$ in some specified way, and then showing that any 
$Y\in W^+$ such that $D^{(2)} Y =0$ and such that 
$\sigma^{(2)}(Y)$ is a power series 
in $\hbar$, lies in the subalgebra $W$ itself. 
\smallskip

2.) The mappings $\sigma^{(i)}$ may be considered as the
 restriction
of one mapping $\sigma$ from ${\cal A}$ to $\Gamma(E[[\hbar]])$.
This mapping is simply given by setting $y$ to zero, i.e., it 
picks out the part of $\hat{X} \in {\cal A}$  not depending on
 $y$. We will use the notions of \cite{fedosov} and call   
$\sigma(\hat{X})$ the {\em symbol} of $\hat{X} \in {\cal A}$.  

\medskip

From now on we will restrict ourselves to one fixed symplectic 
connection $\partial_s$ on $M$ and will only modify the 
connection $\nabla$ on $V$. We will prove two lemmas which will 
be essential in the following sections. 

\begin{lemma} \label{lem1}
Let $\nabla^{(1)}$ and $\nabla^{(2)}$ be  connections on 
$E[[\hbar]]$, $\partial_s$ a fixed symplectic connection, 
 and let $D^{(1)}, D^{(2)}$ denote the corresponding flat Fedosov 
connections, and $U_{(12)}$ the corresponding automorphism.
Assume that the difference of the two connections $\nabla^{(1)}$
 and $\nabla^{(2)}$ is of order $O(\hbar^m)$ for some $m$. Then:
\[  \sigma(U_{(12)} \diamond  \hat{X} \diamond  U_{(12)}^{-1})
 =  \sigma(\hat{X}) +  O(\hbar ^{(m+1)}) \]
for any $\hat{X} \in {\cal A}$ which is covariantly constant with 
respect to $D^{(1)}$.
 \end{lemma}

\medskip \noindent {\it Proof: } 
 By $(\ref{eq:Del_gam})$, $A_{(12)} $ is contained in 
${\cal A}_{2 m +3}$.
Hence, using  $(\ref{eq:U_conj})$, we may conclude that 
$U_{(12)} \diamond  \hat{X} \diamond  U_{(12)}^{-1} - \hat{X} 
\in {\cal A}_{2 m +1}$,
and the symbols $\sigma(U_{(12)} \diamond  \hat{X} 
\diamond  U_{(12)}^{-1}) ,\sigma(\hat{X})$
 may differ only by a term of order $O(\hbar^{m+1})$.
\hfill Q.E.D. ~~~~

\medskip

If we consider three different connections, $U_{(13)}$ is not 
necessarily the same as the composition 
$U_{(23)} \diamond  U_{(12)}$
of $U_{(12)}$ and $U_{(23)}$, but  $ U_{(13)}^{-1} \diamond  
U_{(23)} \diamond  U_{(12)}$
defines an automorphism of the elements in ${\cal A}$ covariantly 
constant with respect to $D^{(1)}$. Although this automorphism 
in general is not simply the identity, we have the following lemma:

\begin{lemma} \label{lem2}
Let $\nabla^{(1)}$, $\nabla^{(2)}$, and $\nabla^{(3)}$ be  
connections on $E[[\hbar]]$,  $\partial_s$ a fixed symplectic connection,
 and let $D^{(i)}$ denote the corresponding flat Fedosov 
connections and $U_{(ik)}$ the corresponding automorphisms of 
${\cal A}$  mapping covariant 
constant sections with respect to $D^{(i)}$ to covariant constant
 sections with respect to $D^{(k)}$.
Assume that the difference of the two 
 connections $\nabla^{(2)}$ and $\nabla^{(3)}$ is of order 
$O(\hbar^m)$ for some $m$. Then:
\[  \sigma(U_{(13)} \diamond  \hat{X} \diamond  U_{(13)}^{-1})
 = \sigma ((U_{(23)} \diamond  U_{(12)}) \diamond  \hat{X} \diamond 
 (U_{(23)} \diamond  U_{(12)})^{-1} )   + O(\hbar ^{(m+2)}) \] 
for any $\hat{X} \in {\cal A}$ which is covariantly constant with 
respect to $D^{(1)}$
 \end{lemma}

\noindent {\em Proof:}  
By the Baker-Campbell-Hausdorff formula, $U_{(23)} \diamond 
 U_{(12)}$ may be 
written as $\exp_\diamond (\frac{i}{\hbar} A) $ with 
$A = A_{(12)} + A_{(23)} + C$, where $C$  is the sum of 
 multiple commutators of 
$\frac{1}{\hbar} A_{(12)}$ and   $\frac{1}{\hbar} A_{(23)}$
multiplied by $\hbar$. From formulas $(\ref{eq:A_def})$ and
$(\ref{eq:Del_gam})$ it is obvious that $A$ does not contain
negative powers of $\hbar$. 

Now, by assumption and $(\ref{eq:A_def}), (\ref{eq:Del_gam})$,
$A_{(23)} $ is contained in ${\cal A}_{2 m +3}$, and is of the
 special form: 
\[ A_{23} = \hbar^{m+1} T^1_{(23)} + \hbar^m T^2_{(23)} +
 T^3_{(23)} \]
where $T^1_{(23)}$ contains only terms at least of degree 1 in the 
fiber coordinate $y$, $T^2_{(23)}$ contains 
only terms at least of degree 3 in $y$, and $T^3_{(23)} \in
 {\cal A}_{2m+4}$. 

Similarly, $ A_{12} = \hbar  T^1_{12} + T^2_{12} + T^3_{12}$ 
with  $T^1_{(12)}$ containing only terms at least of degree 1 in 
the  fiber coordinate $y$, $T^2_{(12)}$ containing 
only terms at least of degree 3 in $y$, and
 $T^3_{(12)} \in {\cal A}_{4}$. Now, since all other terms in the
 Baker-Cambell-Hausdorff formula will be contained in 
${\cal A}_{2m+5}$, 
we may conclude that 
\[C=\frac{1}{\hbar} [\hbar  T^1_{12} + T^2_{12} , \hbar^{m+1}
 T^1_{12} +\hbar^m T^2_{12} ]
 + ...,\] 
where the missing terms 
denoted by dots are in ${\cal A}_{2m+5}$. An inspection of each
 term, using the fact that $T^2_{12},T^2_{23}$ are scalar,  
finally shows that $C$ is of the form:
\[ C = \hbar^m C^1 + \hbar^{m+1} C^2 + \hbar^{m+2} C^3 + C^4, \] 
 where $C^1$ only conatains terms at least quadratic in $y$, $C^2$
 only terms at least of degree 1 in $y$, and 
$C^4 \in {\cal A}_{2m+5}$.
Since $C^1$ and $C^2$ contain no terms independent of $y$, 
they may  contribute to the symbol of 
$ \sigma ((U_{(23)} \diamond  U_{(12)}) \diamond  \hat{X} 
\diamond  (U_{(23)} \diamond  U_{(12)})^{-1}$ 
only through higher order terms of the Moyal product. Hence, 
the whole contribution of $C$ to 
the symbol is of order $\hbar^{m+2}$.

On the other hand, we see from $(\ref{eq:A_def})$ and 
$(\Delta \gamma)_{(13)} =(\Delta \gamma)_{(12)} +
 (\Delta \gamma)_{(23)} $  that 
$A_{(13)} =    A_{(12)} + A_{(13)} +  \tilde{C}$,
 where $\tilde{C} \in  {\cal A}_{2 m +4}$ as well. 
A close inspection of the possible term in $\tilde{C}$ shows again
that it contributes to the symbol of
$  U_{(13)} \diamond  \hat{X} \diamond  (U_{(13)})^{-1}$ only at order
$\hbar^{m+2}$. Thus, the difference of the two symbols is of order
$\hbar^{m+2}$ \hfill Q.E.D.~~~~
 
Let $\tilde{\nabla}$ be an arbitrary (possibly $\hbar$-dependent)
connection on $V[[\hbar]]$, and $\pi_\alpha$ a set of projection
valued sections in $E[[\hbar]]$ with $\pi_\alpha \pi_\beta =
\delta_{\alpha \beta} \pi_\alpha$, $\sum_\alpha \pi_\alpha = \mbox{id}$. 
Define $\nabla = \sum_\alpha 
\pi_\alpha \circ \tilde{\nabla} \circ \pi_\alpha$ as a 
``Berry type connection''. Then $ \nabla \pi_\alpha = 0$, and 
we have the following lemma: 

\begin{lemma} \label{lem3}
With the notations above, and a fixed symplectic connection 
$\partial_s$, let $*$ denote the Fedosov  star product
 corresponding to $\nabla$. Then, 
$\pi_\alpha$ are ``quantum projection'', i.e., 
$\pi_\alpha * \pi_\beta =
\delta_{\alpha \beta} \pi_\alpha $.
\end{lemma}
  
\medskip \noindent {\it Proof: } 
 We will show: $\pi_\alpha * \pi_\beta =\pi_\alpha\cdot \pi_\beta$.
We first note that, since $\nabla \pi_\alpha = 0 ~\forall \alpha$, 
it follows from $(\ref{eq:r_def})$ that $[r,\pi_\alpha] = 0 ~ 
\forall \alpha$ (Since $\pi_\alpha$ does not depend on $y$, there 
is no difference between $\diamond$-commutator
 and ordinary matrix commutator.)
 Then, from $(\ref{eq:cov_const})$: ~~ ${\cal Q}(\pi_\alpha) 
= \pi_\alpha$ for all $\alpha$. But then we have:
 $ \pi_\alpha * \pi_\beta =\pi_\alpha\cdot \pi_\beta= 
\delta_{\alpha \beta} \pi_\alpha $. \hfill Q.E.D. ~~~~

\section{Berry connections and ``block diagonalization'' } 
\label{sec:block_diag}

After the technical preliminaries of the last section, we are now 
in the position to address the actual problem of this paper. 

We consider a Hamiltonian which is a section in the bundle 
$E[[\hbar]]$: \[ H = H_0 + \hbar H_1 + \ldots \] 
We assume that the eigenvalues  $~\lambda_\alpha(x)$ of $H_0(x) 
~(\alpha=1,\ldots r)$, 
have constant  multiplicities $m_\alpha$ throughout phase space. 
We denote by $\pi^{(0)}_\alpha(x)$ the projections of $V_x$  
onto the 
$ i$-th eigenspace of $H_0(x)$. Under the regularity assumptions 
above, these 
eigenspaces form smooth vector bundles over $M$. 

We assume that we are given a hermitian connection on $V$ and a symplectic 
 connection on $M$, which together
define a star product; in many applications there is a 
distinguished choice
of connections and quantization.  For example, in the case of the
Born-Oppenheimer approximation, one may start with a symplectic 
manifold which is
just the nuclear phase space $\real^{2 N}$ where $N$ is three 
times the number of nuclei in the molecule. 
In this case, there is a distinguished quantization
 prescription, namely Weyl  ordering, which corresponds to the
 choice of the trivial connection on the trivial bundle $V$.

Hence, we assume that we are given a definite quantization.
As we allow the symbols of the observables to depend on $\hbar$,
one may alternatively interpret this as the choice 
of a definite symbol calculus. 
 
Nevertheless, we still have the freedom of changing the 
connection defining our star product without changing the physics,
if we apply the corresponding isomorphism of the resulting algebras
at the same time, i.e., we have to correct the symbol $H$ by
applying the isomorphism $\phi$ introduced above. 
We can use this freedom to reduce the problem 
to a simpler one, which is closer to the scalar case. That this 
is possible indeed is the content of the next theorem:

\begin{theorem} \label{thm1}
There exists a formally orthogonal decomposition 
$V[[\hbar]] = \oplus_{\alpha=1}^r V_\alpha$, 
$\dim(V_\alpha)= m_\alpha$ with 
corresponding quantum projections   $\pi^{(\infty)}_\alpha$
(i.e., $\pi^{(\infty)}_\alpha *_\infty \pi^{(\infty)}_\beta=
\delta_{\alpha \beta} \pi^{(\infty)}_\alpha$), and a 
compatible connection $\nabla^{(\infty)}$ 
(i.e. $\nabla^{(\infty)} \pi^{(\infty)}_\alpha = 0 ~ 
\forall \alpha$) such that the corrected symbol $H_{\mbox{corr}}$
preserves the  decomposition:  
$ [ H_{\mbox{corr}}, \pi_\alpha^{(\infty)}] = 0 $
for $\alpha =1, \ldots ,r$. (Here, $[,]$ denotes the $*_\infty$ 
commutator). 
\end{theorem}

\noindent {\it Proof: ~} We show the theorem by induction on 
the order in $\hbar$.

For the connections $\nabla^{(k)}$ to be constructed below 
we denote by $\phi_k$ the isomorphism of the algebras
 $\Gamma(E[[\hbar]])$ with star products $*,*_k$
 constructed from $\nabla$ and $\nabla^{(k)}$,
 respectively:
$~~\phi_k( X * Y) = \phi_k(X) *_k \phi_k(Y) $.

 Let $\pi^{(1)}_\alpha(x) \defeq \pi^{(0)}_\alpha(x)  $ denote
 the projection onto the $i$-th 
eigenspace of $H_0(x)$ (i.e., of the zeroth order term of the 
original Hamiltonian $H$), and define $\nabla^{(1)}$ as the 
Berry-type connection: \[ \nabla^{(1)} \psi \defeq \sum_\alpha 
\pi^{(1)}_\alpha \nabla(\pi^{(1)}_\alpha \psi) ~~.\]
 The difference between the corrected Hamiltonian
 $H^{(1)} = \phi_1(H)$, which is obtained 
from $H$ by applying the isomorphism corresponding to the 
change of the connection from $\nabla$ to $\nabla^{(1)}$ is of 
order $O(\hbar)$ by Lemma \ref{lem1}. 
Hence,  $\nabla^{(1)} \pi^{(1)}_\alpha \equiv 0$
by construction, and $[H^{(1)},\pi^{(1)}_\alpha] =O(\hbar) $ for all $\alpha$.

Now assume that we have found (formally orthogonal) projections 
$\pi^{(k)}_\alpha$ with $\pi^{(k)}_\alpha = \pi^{(0)}_\alpha + 
O(\hbar)$ 
with corresponding  Berry-type connection 
 \[ \nabla^{(k)} \psi  \defeq \sum_\alpha \pi^{(k)}_\alpha
 \nabla(\pi^{(k)}_\alpha \psi)  ~~\]
 such that  the corrected Hamiltonian $H^{(k)}= \phi_k(H)$,
  satisfies $ [H^{(k)},\pi^{(k)}_\alpha ] = O(\hbar^k). $
 We will construct  new orthogonal projections 
$\pi^{(k+1)}_\alpha$ satifying  the requirement:
 \begin{equation} \label{eq:H_comm}  
[H^{(k)}, \pi^{(k+1)}_\alpha] = O(\hbar^{k+1}) ~~ \forall \alpha
\end{equation} 
To this end, we make the ansatz: 
\begin{equation}
\pi^{(k+1)}_\alpha= \exp(i \hbar^{k} A)\pi^{(k)}_\alpha 
\exp(-i \hbar^{k} A) 
\label{eq:pi_new}
\end{equation}
for a fixed $A$  independent of $\alpha$.
 If we can find such an $A$, equation 
$(\ref{eq:pi_new})$ automatically defines new formally 
orthogonal projections (i.e. $\pi^{(k+1)}_\alpha 
*_{k+1}\pi^{(k+1)}_\beta  
= \pi^{(k+1)}_\alpha \pi^{(k+1)}_\beta =
\delta_{\alpha \beta} \pi^{(k+1)}_\alpha $, according to 
lemma \ref{lem3}.) With this ansatz $\pi^{(k+1)}_\alpha$ 
is given up   to order $O(\hbar^{k+1})$ by 
\[  \pi^{(k+1)}_\alpha= \pi^{(k)}_\alpha  + i \hbar^{k} 
                      [A,\pi^{(k)}_\alpha ] + O(\hbar^{k+1}) \]

Hence,  equation $(\ref{eq:H_comm})$ yields up  to order
 $O(\hbar^{k+1})$
the condition:
\[ [H^{(k)}, \pi^{(k)}_\alpha ] +i \hbar^{k} 
 [H_0,[ A,\pi^{(0)}_\alpha ]] = O(\hbar^{k+1}) 
~~ \forall \alpha\]
As the $\pi^{(k)}_\alpha $ are projections, we may write: 
\[H^{(k)} = \sum_\alpha \pi^{(k)}_\alpha H^{(k)}
 \pi^{(k)}_\alpha  + \hbar^{k} W\]
where the first term on the right hand side commutes with
 $\pi^{(k)}_\alpha$ and 
\[\hbar^{k} W = \sum_{\beta  \not= \gamma} \pi^{(k)}_\beta H^{(k)}
 \pi^{(k)}_\gamma
 = \sum_{ \beta \not= \gamma} \pi^{(k)}_\beta 
( [ H^{(k)}, \pi^{(k)}_\gamma ]
  \pi^{(k)}_\gamma . \]

As the commutator on the right hand side is of order 
$O( \hbar^{k})$  
the left hand side is of the same order. Hence, we have to solve:
\[ \hbar^{k} [W , \pi^{(0)}_\alpha ] +i \hbar^{k} 
 [H_0,[ A,\pi^{(0)}_\alpha ]] = O(\hbar^{k+1}) 
~~ \forall \alpha \] 
Note that in this equation only the projections $\pi^{(0)}_\alpha$ 
~($\alpha = 1,\ldots, r)$ appear. 
A solution to this equation is given by:
\[ A =  i \sum _{\alpha \not= \beta }
 \frac{ \pi^{(k)}_\alpha  W \pi^{(k)}_\beta}
{\lambda_\alpha - \lambda_\beta} ,  \] 
as  may be easily verified using 
$\pi^{k}_\alpha W \pi^{k}_\alpha = 0$
for all $\alpha$ by construction of $W$.

Using the new projections we may define a 
new Berry-type connection 
 $\nabla^{(k+1)}$ which preserves the decomposition defined by the 
projections even to arbitrary order in $\hbar$. As the difference 
of the projections is of order $O(\hbar^{k})$, 
the difference of the 
connections $\nabla^{(k)}$ and $\nabla^{(k+1)}$   is of the 
same order, 
and hence, by Lemmas \ref{lem1} \ref{lem2}, the corrected symbol 
$H^{(k+1)}= \phi_{k+1}(H) $ differs from $H^{(k)}$ 
only at order  $O(\hbar^{k+1})$
 Thus, by  $(\ref{eq:H_comm})$, ~$ H^{(k+1)}$
is compatible with the decomposition up to $O(\hbar^{k+1})$:
$  [H^{(k+1)},\pi^{(k+1)}_\alpha ] = O(\hbar^{k+1}) $

As $\nabla^{(k)}$ and $\nabla^{(k+1)}$ differ only at 
order $O(\hbar^k)$,
the connections $\nabla^{(k)}$ have a well defined 
limit $\nabla^{\infty}$,
which by induction fulfills the assertions of the theorem.
\hfill Q.E.D.~~~~

\bigskip

The importance of theorem \ref{thm1} lies in the following 
two corollaries:
\begin{corollary}
Let $X,Y \in \Gamma{E[[\hbar]]}$ such that 
$[X, \pi^{(\infty)}_\alpha ]
= [Y, \pi^{(\infty)}_\alpha ] = 0$ for all $\alpha$.
 Let $*_\infty$ denote
the star product constructed from $\nabla^{(\infty)}$ by Fedosov's
construction. Then $[ X*_\infty Y, \pi^{(\infty)}_\alpha ] = 0 $ 
for all $\alpha$.
\end{corollary} 

\noindent {\em Proof:} 
As $\nabla^{(\infty)}  \pi^{(\infty)}_\alpha  = 0$, the curvature
 of $\nabla^{(\infty)} $ commutes with $\pi^{(\infty)}_\alpha$. As 
$\delta ^{-1} $ only acts on the second factor of 
${\cal A} = \Gamma( E[[\hbar]] \tensor W)$ and the Moyal product
$\diamond$ only involves derivatives with respect to the 
fibre coordinates $y$, we may conclude from $(\ref{eq:r_def})$ 
by induction on the total degree that $r$ commutes with 
$\pi^{(\infty)}_\alpha  $.
 (Observe that we may write the term
$r^2 = r \diamond r $ as $1/2 [ r,r]_\diamond $). 

Now, the covariant constant continuation 
$\hat{X} ={\cal Q}^{(\infty)} (X)$ of
$X$ is given by \cite{fedosov}:
\begin{equation} \label{eq:X_cov}
 \hat{X} = X + \delta^{-1} \left( \partial \hat{X} +
   \left[\frac{i}{\hbar} r, \hat{X}\right]_\diamond \right) 
\end{equation}
and we may again conclude inductively:
$[\hat{X}, \pi^{(\infty)}_\alpha ] = 0$, 
and similarly for $\hat{Y}$.

Using again the fact that the Moyal product $\diamond $ on 
$\cal A$ 
only involves derivatives with respect to $y$,
it follows  that $ \hat{X} \diamond \hat{Y}$ commutes with
$\pi^{(\infty)}_\alpha $ for all $\alpha$. Hence, 
by taking the symbol,
i.e., by setting $y=0$, we get the assertion of the corollary.
\hfill Q.E.D. ~~~~~

\begin{corollary}
Denote by $\phi_\infty$ the isomorphism of the 
algebras $\Gamma(E[[\hbar]])$
with products $*$ and $*_\infty$. Let 
\[ {\cal O} \defeq \{  \phi_\infty^{-1}(\tilde{X}) |\;
 \tilde{X} \in \Gamma(E[[\hbar]]), \;
 [\tilde{X}, \pi^{(\infty)}_\alpha ] = 0 ~\forall \alpha
 \}. \]   
Then, $ {\cal O} $ is a subalgebra of $ \Gamma(E[[\hbar]])$ 
for which the Heisenberg time evolution 
\[ \dot{X} = \frac{1}{i \hbar} [ H, X ]_* \]
is a well defined differential equation.
\end{corollary} 

\noindent {\em Proof: }  
$ {\cal O} $ is a subalgebra of $\Gamma(E[[\hbar]]) $   by the
 preceding corollary.  We only have to show that 
$[ H,X]_* = O(\hbar)$, since then
there will be no negative powers of $\hbar $ in the Heisenberg time 
evolution. 

Denoting again $\phi_\infty(H) $ by $H_{corr}$, we know 
that $H_{corr}$ is of the form $\sum_\alpha \lambda_\alpha 
\pi^{(\infty)}_\alpha  + O(\hbar)$.  

Now, by $(\ref{eq:r_def})$, $r$ is of the form 
$ r = 1 \tensor \hat{r} + O(\hbar)$ 
for a scalar one form $\hat{r}$. From this one may conclude that 
the covariantly constant continuation 
${\cal Q}^{(\infty)}(H_{corr})$ of $H_{corr}$  is of the form 
 $\sum_\alpha \rho_\alpha \pi^{(\infty)}_\alpha  + O(\hbar)$
for some scalar $\rho_\alpha $. Thus, since $\tilde{X}$
commutes with $\pi_\alpha^{(\infty)}$, we may conclude that  
 the classical matrix commutator does not contribute to 
$[H_{corr}, \tilde{X} ]_{*_\infty}$ at zeroth order in $\hbar$. 
Hence, the commutator is of order $\hbar$.
Applying $\phi_{\infty}^{-1}$, the corollary follows.
~~~~ \hfill Q.E.D. ~~~~~

\section{Correction at order $\hbar$}

 \label{sec:corrections}
In the previous section we have achieved a complete reduction 
of the multicomponent WKB problem to a set of different 
polarizations. 
In this section, we will explicitly compute the 
corrections to the symbol of 
the Hamiltonian at order $\hbar$. Thus, this section 
will generalize the results of \cite{li-fl:geometric},
 and give a completely geometric derivation
of the results derived there for trivial bundles only.
 We achieve this by first using the Berry type connection 
as in the prove of theorem \ref{thm1} in the previous section.
 However, the connection is not uniquely defined 
by the compatibility conditions:
Hence, in the second subsection, we will investigate the 
dependence of the structure of this phase  
 on the choice of a compatible connection (in the sense of theorem \ref{thm1}).
 It will turn out that it 
is unique up to a Berry type phase, and that 
the Berry type connection
considered so far is geometrically distinguished 
among all compatible connection.  

\subsection{ Berry type connection}
To compute the correction of the symbol $H$ at order $\hbar$, 
we only have to carry out  the first two steps of the iterative
 procedure in the proof of the previous theorem. 
 Actually, it is not necessary 
to compute $H^{(2)}$, but it is sufficient to compute the ``block
diagonal'' part of $H^{(1)}$ since 
\begin{eqnarray*}
H_{corr} &=& H^{(2)} +O(\hbar^2) = 
\sum_\alpha \pi^{(2)}_\alpha  H^{(1)} 
\pi^{(2)}_\alpha + O(\hbar^2)\\ &=&
 \sum_\alpha \exp(i \hbar A) \pi^{(1)}_\alpha 
  \exp(-i \hbar A)  H^{(1)}   
 \exp(i \hbar A) \pi^{(1)}_\alpha  \exp(-i \hbar A) \\ 
& = &\sum_\alpha \exp(i \hbar A) \pi^{(1)}_\alpha   (  H^{(1)}   
 + i \hbar [H^{(1)},A] )\pi^{(1)}_\alpha
     \exp(-i \hbar A) + O(\hbar^2)\\ 
 & = & \sum_\alpha \exp(i \hbar A) \pi^{(0)}_\alpha
         H^{(1)} \pi^{(0)}_\alpha  
\exp(-i \hbar A) + O(\hbar^2)
\end{eqnarray*}
with $A$ as defined in the proof of Theorem \ref{thm1}, 
and $\pi^{(1)}_\alpha = \pi^{(0)}_\alpha$.

Here, we used the fact that zeroth degree part $H^{(1)}_0$
of $H^{(1)}$ is just $H^{(0)}_0$, and 
that $\pi^{(0)}$ is the projection 
onto the eigenspaces of $H^{(0)}_0$.  
Hence, in order to determine the correction of the symbol at order 
$\hbar$ up to a unitary rotation, we only have 
to compute  $\pi^{(0)}_\alpha  H^{(1)} \pi^{(0)}_\alpha $
for all $\alpha$. 

Defining $\gamma_l \defeq \sum_\alpha \pi^{(0)}_\alpha
 (\nabla_l \pi^{(0)}_\alpha )$
we get by  formulas $(\ref{eq:U_conj}), (\ref{eq:A_def})$:
\begin{eqnarray}
\pi^{(0)}_\alpha  H^{(1)}\pi^{(0)}_\alpha & =& 
\pi^{(0)}_\alpha \Bigl( H^{(0)} + \sigma( [\gamma_l y^l , 
\nabla_\alpha H^{(0)}_0 y^i ]_\diamond  )\nonumber \\[1mm]&&
 + \frac{1}{2}  \sigma( [\gamma_l y^l , 
[\gamma_m y^m , H^{(0)}_0 ]_{\mbox{\scriptsize cl}} 
]_\diamond  \Bigr) \pi^{(0)}_\alpha   + O(\hbar^2)\nonumber \\[1mm]
& = &   \pi^{(0)}_\alpha  H^{(0)}\pi^{(0)}_\alpha 
 + \pi^{(0)}_\alpha \left( \omega^{il}
\frac{i \hbar}{2} \left( \gamma_i \nabla_l H^{(0)}_0  
- \nabla_i H^{(0)}_0  \gamma_l
\right) \right) \pi^{(0)}_\alpha 
\nonumber \\[1mm] && + \pi^{(0)}_\alpha \frac{i \hbar}{4} 
\omega^{il}
\left(  \gamma_i  
[\gamma_l  ,  H^{(0)}_0 ]_{\mbox{\scriptsize cl}} 
-    
[\gamma_i  ,  H^{(0)}_0 ]_{\mbox{\scriptsize cl}} \gamma_l \right) 
\pi^{(0)}_\alpha  + O(\hbar^2) \label{eq:H1}
\end{eqnarray}
Here, $[\cdot,\cdot]_{\mbox{\scriptsize cl}} $ 
denotes the usual matrix
commutator, and $[ \cdot, \cdot]_\diamond $ the Moyal 
product commutator.

In the computation, the following relations (following trivially 
from the projection property) prove to be useful:
\[  \pi^{(0)}_\alpha (\nabla \pi^{(0)}_\beta) \pi^{(0)}_\alpha 
 =0 ~~~\forall \alpha,\beta \] 
\[   \pi^{(0)}_\alpha (\nabla \pi^{(0)}_\beta) \pi^{(0)}_\beta=
     \pi^{(0)}_\alpha (\nabla \pi^{(0)}_\beta), ~~~
\pi^{(0)}_\beta (\nabla \pi^{(0)}_\beta) \pi^{(0)}_\alpha 
(\nabla \pi^{(0)}_\beta) \pi^{(0)}_\alpha ~~~ \forall \alpha \not=
 \beta  \] 

Using $ H^{(0)}_0(x)  =\sum_\alpha  \lambda_\alpha(x) 
\pi^{(0)}_\alpha(x)$, we get
for the second term on the right hand side of the last equation in 
$(\ref{eq:H1})$:
\[
\begin{array}{l}
 \frac{i \hbar }{2} \omega^{il}
 \Bigl( \sum \limits_\beta \pi^{(0)}_\alpha
(\nabla_i \pi^{(0)}_\alpha ) \nabla_l
 ( \lambda_\beta \pi^{(0)}_\beta )\pi^{(0)}_\alpha
- \sum \limits_{\beta,\gamma}
 \pi^{(0)}_\alpha \nabla_i (\lambda_\beta \pi^{(0)}_\beta)
\pi^{(0)}_\gamma (\nabla_l \pi^{(0)}_\gamma)
 \pi^{(0)}_\alpha \Bigr)   \\[4mm]
=  \frac{i \hbar }{2} \omega^{il} 
\Bigl( \lambda_\alpha \pi^{(0)}_\alpha
(\nabla_i \pi^{(0)}_\alpha ) (\nabla_l \pi^{(0)}_\alpha) 
- \sum \limits_{\beta} \lambda_\beta \pi^{(0)}_\alpha 
       (\nabla_i \pi^{(0)}_\alpha)
\pi^{(0)}_\beta  (\nabla_l \pi^{(0)}_\alpha) \Bigr. \\[4mm] \Bigl. 
\qquad - \sum \limits_{\beta} 
\lambda_\beta \pi^{(0)}_\alpha (\nabla_i \pi^{(0)}_\beta)
\pi^{(0)}_\alpha  (\nabla_l \pi^{(0)}_\alpha)
+ \sum \limits_{\beta, \gamma} \lambda_\beta 
\pi^{(0)}_\alpha \nabla_i ( \pi^{(0)}_\beta)
\pi^{(0)}_\gamma (\nabla_l \pi^{(0)}_\alpha) \Bigr) \\[4mm]
= i \hbar \omega^{il} \sum \limits_{\beta}
 \lambda_\beta \pi^{(0)}_\alpha (\nabla_i \pi^{(0)}_\beta)
 (\nabla_l \pi^{(0)}_\alpha)
\end{array}\] 

For the last term in $(\ref{eq:H1})$ we get:
\[\begin{array}{l}  
 \frac{i \hbar}{4} \omega^{il}
\Bigl(  \sum_{\beta,\gamma } \pi^{(0)}_\alpha 
(\nabla_i \pi^{(0)}_\alpha)
  [ \pi^{(0)}_\beta \nabla_l \pi^{(0)}_\beta,
 \lambda_\gamma \pi^{(0)}_\gamma ]_{\mbox{\scriptsize cl}} 
\pi^{(0)}_\alpha
\\[4mm]    
 \qquad \qquad  - \sum \limits_{\beta , \gamma, \rho }  
\pi^{(0)}_\alpha 
  [ \pi^{(0)}_\beta (\nabla_i \pi^{(0)}_\beta), 
\lambda_\gamma \pi^{(0)}_\gamma ]_{\mbox{\scriptsize cl}} 
 \pi^{(0)}_\rho (\nabla_l \pi^{(0)}_\rho)  \pi^{(0)}_\alpha \Bigr)
 \\[4mm]   
=  \frac{i \hbar}{4} \omega^{il}
\Bigl( \lambda_\alpha \sum \limits_{\beta} \pi^{(0)}_\alpha
 (\nabla_i \pi^{(0)}_\alpha)
   \pi^{(0)}_\beta (\nabla_l \pi^{(0)}_\beta) \pi^{(0)}_\alpha
- \sum \limits_{ \gamma }  \lambda_\gamma  \pi^{(0)}_\alpha 
 (\nabla_i \pi^{(0)}_\alpha)
   \pi^{(0)}_\gamma (\nabla_l \pi^{(0)}_\gamma ) \pi^{(0)}_\alpha
\\[4mm]    \qquad 
- \sum \limits_{ \gamma }  \lambda_\gamma  \pi^{(0)}_\alpha 
 (\nabla_i \pi^{(0)}_\alpha)
   \pi^{(0)}_\gamma (\nabla_l \pi^{(0)}_\gamma ) \pi^{(0)}_\alpha
 + \sum \limits_\rho 
\lambda_\alpha  \pi^{(0)}_\alpha (\nabla_i \pi^{(0)}_\alpha)
   \pi^{(0)}_\rho (\nabla_l \pi^{(0)}_\rho) 
    \pi^{(0)}_\alpha \Bigr)\\[4mm]   
= - \frac{i \hbar }{2} \omega^{il}  
\sum \limits_\gamma \lambda_\gamma  \pi^{(0)}_\alpha
(\nabla_i \pi^{(0)}_\gamma)(\nabla_l \pi^{(0)}_\alpha)
\end{array} \]
Thus, we finally get:
\begin{eqnarray*}
\pi^{(0)}_\alpha  H^{(1)}\pi^{(0)}_\alpha &=& \pi^{(0)}_\alpha 
        H^{(0)}\pi^{(0)}_\alpha
+  \frac{i \hbar }{2} \omega^{il}  
       \sum \limits_\gamma \lambda_\gamma  \pi^{(0)}_\alpha
(\nabla_i \pi^{(0)}_\gamma)(\nabla_l \pi^{(0)}_\alpha)
 \\ & = & 
\pi^{(0)}_\alpha  H^{(0)}\pi^{(0)}_\alpha
+  \frac{i \hbar }{2} \omega^{il}   \lambda_\alpha
         \pi^{(0)}_\alpha
(\nabla_i \pi^{(0)}_\alpha)(\nabla_l \pi^{(0)}_\alpha) \\&&
+  \frac{i \hbar }{2} \omega^{il}  
       \sum \limits_{\gamma \not= \alpha}
 \lambda_\gamma  \pi^{(0)}_\alpha 
(\nabla_i \pi^{(0)}_\gamma) \pi^{(0)}_\gamma 
       (\nabla_l \pi^{(0)}_\alpha)
\end{eqnarray*}

In order to give a geometric interpretation
 of the correction terms 
we compute the curvature of the Berry connection:
\begin{eqnarray*} 
 F^B \psi &=& \sum_\alpha \pi^{(0)}_\alpha \circ  \nabla
\circ \pi^{(0)}_\circ  \nabla \pi^{(0)}_\alpha \psi 
= \sum_\alpha  \pi^{(0)}_\alpha \nabla ^2 \pi^{(0)}_\alpha
 + \sum_\alpha  \pi^{(0)}_\alpha (\nabla \pi^{(0)}_\alpha) 
       (\nabla \pi^{(0)}_\alpha)
\\[2mm] & =&  \sum_\alpha \pi^{(0)}_\alpha F^{o} \pi^{(0)}_\alpha 
+ \sum \pi^{(0)}_\alpha(\nabla \pi^{(0)}_\alpha)
\wedge (\nabla \pi^{(0)}_\alpha) 
\end{eqnarray*}
where $F^o$ denotes the curvature of 
the original connection $\nabla$,
and the covariant derivatives have to be interpreted as covariant 
exterior derivatives, acting on forms.

Hence, the  coefficient of $\lambda_\alpha$ of the $\alpha$-th 
block in the correction at order $\hbar$ is 
$ \omega^{-1} ( \pi^{(0)}_\alpha F^B \pi^{(0)}_\alpha - 
\pi^{(0)}_\alpha F^0 \pi^{(0)}_\alpha)$.
 Here, $\omega^{-1}$ is the Poisson
tensor. The curvature of the Berry connection  is block diagonal
by construction, the effect of the projection operators on both 
sides is just to pick out the right block, corresponding to 
$\alpha$. In particular, if the original connection is flat, the
coefficient of $\lambda_\alpha$ on the $\alpha$-th block  
 of the correction is exactly the Poisson  curvature of 
the Berry connection, i.e., the contraction of the curvature with
the Poisson tensor.

In order to interpret the remaining terms we define a
vector bundle  analog of the second 
fundamental form for embedded submanifolds:

For arbitrary $\beta \not= \alpha$ set
\[ \begin{array}{lll}  
\tilde{S}^{\beta \alpha} : & \Gamma(V^{(1)}_\alpha) &
 \rightarrow  \Gamma(V^{(1)}_\beta)
\tensor \Gamma(\Lambda^{1}(M)) \\
& \psi & \mapsto \pi^{(1)}_\beta \circ  \nabla \circ 
 \pi^{(1)}_\alpha 
\psi = \pi^{(1)}_\beta (\nabla \pi^{(1)}_\alpha) \psi
\end{array}\]
$\tilde{S}^{\beta \alpha}$  obviously is tensorial for   
$\beta \not= \alpha$, hence it induces a mapping
$ S^{\alpha \beta}: V^{(1)}_\alpha  \rightarrow
  V^{(1)}_\beta \tensor 
\Lambda^{1}(M)$
which is a vector bundle  analog of the second fundamental form:
It measures the extent to which parallel transport with respect
to $\nabla$ tends to rotate a vector from a given eigenspace into
the others. 

For fixed $\alpha \not= \beta$, we may form 
$S^{\beta \alpha} \wedge S^{\alpha \beta}$, 
which is a two-form with values in the endomorphims 
of $V_\alpha$, and hence may be contracted with the Poisson tensor
to define an endomorphism of $V_\alpha$, which is just the 
coefficient of $\lambda_\beta$ in the $\alpha$-th block 
of the first order correction of $H$.

\subsection{Arbitrary compatible connection}
For an arbitrary  connection $\tilde{\nabla}$  
which is compatible in the sense of Theorem \ref{thm1} up to
some order $\hbar^k$, we first consider the  correction at 
order $\hbar$. By Lemma \ref{lem2}, we may compute the correction
at order $\hbar$ by first going over to the Berry connection, and
then in a second step to the connection $\tilde{\nabla}$, as the 
the corrected symbol obtained in this way
 differs from he one obtained
by going over directly from $\nabla$ to $\tilde{\nabla}$ is 
of order $O(\hbar^2)$.

As both $\nabla^{(1)}$ and $\tilde{\nabla}$ are compatible with 
the projections $\pi^{(0)}_\alpha$   up to $O(\hbar)$, 
the difference
$\Delta \gamma$ of the respective connection forms commutes with 
$\pi^{(0)}_\alpha$ and $H^{(0)}$ up to order $O(\hbar)$.
 Hence, denoting the corrected symbol corresponding to the new
connection $\tilde{\nabla}$ by $\tilde{H}$ we get by
formulas $(\ref{eq:U_conj})$ and $(\ref{eq:A_def})$  :
\begin{eqnarray*}
\pi^{(0)}_\alpha \tilde{H}\pi^{(0)}_\alpha &\!\!=\!\!&
\pi^{(0)}_\alpha \left( H^{(1)} 
+ \sigma( [(\Delta \gamma)_l y^l , 
       \nabla^{(1)}_i H^{(0)}_0 y^i ]_\diamond  )
\right.\\&& \left.+ \frac{1}{2}  
       \sigma( [(\Delta \gamma)_l y^l ,
        [(\Delta \gamma)_m y^m , 
               H^{(0)}_0 ]_{\mbox{\scriptsize cl}} 
]_\diamond  \right) \pi^{(0)}_\alpha  + O(\hbar^2)
\\
& \!\!=\!\! & \pi^{(0)}_\alpha  
       H^{(1)} \pi^{(0)}_\alpha + \pi^{(0)}_\alpha  \omega^{il}
\frac{i \hbar}{2} \left(
       (\Delta \gamma)_i \nabla_l H^{(0)}_0  
       - \nabla_i H^{(0)}_0 (\Delta \gamma)_l
 \right) \pi^{(0)}_\alpha + O(\hbar^2) \\
& \!\!=\!\! & \pi^{(0)}_\alpha  
H^{(1)} \pi^{(0)}_\alpha + \pi^{(0)}_\alpha  \omega^{il}
       {i \hbar} \left((\Delta \gamma)_i  (\partial_l 
       \lambda_\alpha) \right) \pi^{(0)}_\alpha + O(\hbar^2)\\
& \!\!=\!\! & \pi^{(0)}_\alpha  H^{(1)} 
       \pi^{(0)}_\alpha + \pi^{(0)}_\alpha  \omega^{il}
{i \hbar} \left( (\Delta \gamma) (X_{ \lambda_\alpha} \right) 
       \pi^{(0)}_\alpha  + O(\hbar^2)
\end{eqnarray*}
Here, $X_{ \lambda_\alpha} $ denotes the Hamiltonian vector field
corresponding to $\lambda_\alpha$.  

Hence, we see that for an arbitrary connection compatible with 
$\pi^{(0)}_\alpha$ the lowest order corrections differ 
just by a Berry
phase term. The additional geometric term involving 
the eigenvalues 
themselves and not their derivatives are unique and always 
contain the Poisson curvature and the second fundamental form
of the Berry-connection and {\em not} the Poisson curvature 
of the chosen connection. Hence, the Berry connection is 
distinguished among all connections satisfying Theorem \ref{thm1}.

\section{Outlook} \label{sec:scalar} 

In the preceding sections, we have achieved a complete block 
diagonalization of the considered problem: We have found 
a decomposition of $V[[\hbar]]$
into subspaces, such that the symbol of the Hamiltonian 
and the connection 
are compatible with this decomposition. 
Furthermore, we have found a
maximal set of observables which is compatible with 
this decomposition
as well in that sense, that their time evolution  does 
not contain any negative powers of $\hbar$. Hence, for the purposes
of WKB approximation, one may treat each block separately.  
 
However, the problem of multicomponent WKB 
  has not been completely 
reduced to a scalar problem. In this section, we will show that
it can still be reduced  closer to a purely 
scalar one, 
but only at the price of non-canonical choices of a connection. 
Nevertheless, this further reduction proves to be useful for a 
discussion of obstructions to a complete reduction to a scalar 
problem. Furthermore, for this discussion, only parallel transport 
along the Hamiltonian vector field will play a role, which does 
not depend on the choices made in the connection:

We have seen  in the last section that 
the choice of a different connection 
leads at order $\hbar$  to the addition of a Berry-phase term 
$\gamma(X_{\lambda_\alpha})$ to the $\alpha$-th block 
of the corrected symbol. 
We may use the Berry-phase term 
to simplify the problem by choosing the 
connection such that this term just compensates the remaining 
terms at order $\hbar$, so that the Hamiltonian is reduced to a
set of scalar Hamiltonians up to order 
$O(\hbar^2)$ on the $\alpha$th block. This is 
always possible as long as the eigenvalue functions have no
 critical points, where their hamiltonian vector fields vanish. 
In practice this means
that one has to cut out those points. However, this is not a
real problem if one has applications to WKB in mind, 
as  the Lagrangian submanifolds relevant to the WKB approximation 
lie in level sets of the eigenvalue functions  and hence will
generically not
intersect the   critical points. It is possible to iterate 
the procedure above, which gives the following theorem:

\begin{theorem} \label{thm2}
Let $\tilde{M}$ be any open submanifold of $M$ not containing 
critical points of the eigenvalue functions $\lambda_\alpha$.
Denote the embedding of ${\tilde M}$ by $i$, and
let $\tilde{V}$ be the pull back bundle $i^*(V)$, i.e.,
  the subbundle of $V$  obtained 
by restricting $V$ to $\tilde{M}$. 
Then, there is an ($\hbar$ dependent) connection 
$\tilde{\nabla}^{\infty}$
 and a decomposition 
 $\tilde{V}[[\hbar]] = \oplus_{\alpha=1}^m \tilde{V}_\alpha$
  fulfilling Theorem  \ref{thm1} 
such that the corrected  Hamiltonian restricted 
to $\tilde{V}_\alpha$ 
is scalar, namely the eigenvalue function $\lambda_\alpha $ of 
$H^{(0)}$ for each $\alpha$.  
\end{theorem}

As stated before, the connection defined in this way is far from
unique: Only the partial connection for each block, 
giving covariant derivatives 
in the direction of the respective Hamiltonian vector field, 
is uniquely defined
by  the iteration procedure above. However, we may always find a 
continuation
of this partial connection to a connection on the whole bundle.

The result of Theorems \ref{thm1} and \ref{thm2} 
is optimal in the following
sense: We started from a decomposition 
of $V$ into the bundles of eigenvectors
of $H_0$. However, due to the higher order corrections, the 
eigenvalues of the quantum mechanical Hamiltonian operator will 
in general have lower degeneracy than the eigenvalues of the 
principal  symbol $H_0$. 
One might hope to find a finer decomposition of the vector bundle 
$V[[\hbar]]$, so that one ends up with one dimensional subbundles 
and ``formal $U(1)$-connections''. However, this is not possible 
in general, not even at order $O(\hbar^2)$: 
The connection constructed in the previous theorem has in general
a $U(m_1) \times \ldots U(m_r)$ holonomy and cannot be
replaced by a connection with
 a $U(1) \times \ldots \times U(1)$ holonomy without destroying
the block-diagonality of the corrected symbol of the Hamiltonian.

To understand this fact better, we first consider the case of a
two dimensional phase space as base space. In this special case
it is always possible to reduce the problem 
to one-dimensional subbundles,
at least in a tubular neighborhood of any energy 
level set $\lambda_\alpha^{-1}(E)$
for a non-critical value $E$:

To show the existence of such a decomposition and a 
compatible  connection we start with a connection
$\tilde{\nabla}^\infty$
for which the Hamiltonian is scalar on each block, as in theorem
\ref{thm2}. As the connection to be defined will respect the
block structure, we may restrict ourselves to one block 
$\alpha$ and only need to consider the corresponding eigenvalue 
function $\lambda_\alpha$.

If $\lambda_\alpha^{-1}(E)$ is diffeomorphic to $\real$, 
we may choose a transversal manifold $S$ at some point $p \in
 \lambda_\alpha^{-1}(E)$. Choosing  an arbitrary decomposition into
one-dimensional subbundles on $S$, we may define a decomposition
on a whole tubular neighborhood of $\lambda_\alpha^{-1}(E)$
by parallel transport along the  Hamiltonian flow of 
$\lambda_\alpha$.
Denoting by $\tilde{\pi}_k$ the corresponding one-dimensional
projections, we define a new connection $\hat{\nabla}$ as
$\sum_k \pi_k \circ \tilde{\nabla}^\infty \circ \pi_k$.
 Since we have a set of trivializing sections covariantly
 constant along the Hamiltonian
flow both for $\tilde{\nabla}^\infty$ and for $\hat{\nabla}$, it 
follows: 
\[\hat{\nabla}_{X_{\lambda_\alpha}} u 
=\tilde{\nabla}^{\infty}_{X_{\lambda_\alpha}} u
\]
for any section $u$ in the  the $\alpha$-th block, and hence the 
corrected symbol 
is not modified at order $\hbar$ by going over 
to the new connection.
Hence, the Hamiltonian is still scalar (at least up to 
order $O(\hbar^2)$,
and we have found the connection and decomposition we 
are looking for
up to the same order. This procedure can be iterated again,
and we can thus reduce the problem to a purely scalar one.

If  $\lambda_\alpha^{-1}(E)$ is diffeomorphic to $S^1$, then
the situation is a  bit  more involved: 
Due to the periodicity
of the flow, we cannot simply define a decomposition into 
subbundles on a complete
tubular neighborhood of the level set by parallel transport of an 
arbitrary decomposition on a transversal submanifold $S$.
Instead, defining  $U(p)$ as the holonomy for
$\tilde{\nabla}^{\infty }$ along one period of the 
Hamiltonian flow line starting at the point $p\in S$,
we choose the complex eigenspaces of $U(p)$ (which always exist
as $U$ is unitary) on $S$ and parallel transport these along the
Hamiltonian flow of $\lambda_\alpha$.
 (If the eigenspaces
of $U$ are not one-dimensional then we have to choose one
dimensional subspaces in a smooth way along $S$.) We may define
$\hat{\nabla}$ as before and get a  connection with the desired
properties.

We note that we can even modify the connection further 
by adding a $u(1)\times u(1)$-valued one-form to the
 connection form such that the holonomy along the
Hamiltonian flow vanishes and the  symbol of the 
Hamiltonian  is  modified without destroying
diagonality: The corresponding phase is moved  in this way from 
the connection to the  symbol.

These observations explain the special role of two-dimensional 
phase spaces for multicomponent WKB approximation. 
Now, an analogous construction obviously does  not work generally 
if the
underlying phase space is more than two dimensional:
Even if the Hamiltonian system defined by $\lambda_\alpha$ is
integrable, the flow will generically be only quasiperiodic.
 This does not only imply that the above construction does not
work, but even that such a decomposition into one dimensional 
subbundles with a compatible connection and scalar symbols does 
not exist:

Namely, assume such a decomposition would exist. Let $\gamma$
denote the integral curve of $X_{\lambda_\alpha} $ through
a fixed point $p$. 
 Then (pulling back the respective bundles with the inclusion map
$i: \gamma \rightarrow M$)
we first claim that the decomposition of $i^* V[[\hbar]]$
 into one-dimensional
subbundles is uniquely determined by the decomposition at the
point $p$ and the requirement that the corrected symbol
 is diagonal with respect to the decomposition.  
 To see this, we first note that
given any such decomposition with compatible connection, it is determined 
by the decomposition at $P$ through parallel transport. Hence, if there 
are several such decompositions, they must correspond to different
connections. 

However, we may
modify any such connection without changing the decomposition
 by adding a $u(1) \times \ldots \times u(1)$
valued one-form to the connection form. By a proper choice, we may achieve  
that the corrected symbol on the $\alpha$-th block is simply
 $\lambda_\alpha$. 
 Now assume that we are given two different decompositions
with corresponding (properly modified) connections
 which coincide at the point $p$. Then, as the corrected
symbols simply is $\lambda_\alpha$ in both cases, the difference
of the connection forms of the two connections (modified as above)
must vanish on $X_{\lambda_\alpha}$, hence parallel transport along
$\gamma$ coincides for both connections. As the decomposition
has to be invariant under parallel transport, 
the two decomposition must coincide over $\gamma$.

Hence, there is at most one such decomposition, and it determined by
parallel transport with the connection in theorem \ref{thm2}.
Now, due to the quasi-periodicity, the curve $\gamma$ comes
arbitrarily close to $p$ for certain arbitrarily large times $T_i$.
However, there will generically be no one-dimensional subspace
$W_p$ in the fiber $V_p$ over $p$
which, when parallel transported along $\gamma$, will   be
in a suitable sense close to $W_p$ for all the times $T_i$.
 Hence, there cannot  exist  a continous decomposition into
one-dimensional subspaces with the desired properties which is
defined in a whole neighborhood of $\gamma$.

The difference to the situation in perturbation theory comes from
the Berry-phase term. Hence, if one tries to completely diagonalize
the  corrected symbol, one has to solve a differential equation 
along the flow of $\lambda_\alpha$,
 and not just simply to diagonalize
a matrix. Thus, although it is always possible to completely diagonalize 
the symbol locally, this is  not possible  in general globally.


\begin{thebibliography}{99}
\bibitem{ba-we:lectures}
Bates,~S.M., and Weinstein,~A.,  Lectures on the Geometry of Quantization,
{\em Berkeley Mathematics Lecture Notes }
\bibitem {bayen}
        {F. Bayen, M. Flato, C. Fronsdal, A. Lichnerowicz,
           D. Sternheimer:}
        {Deformation Theory and Quantization.}
        {\it Annals of Physics} {\bf 111} (1978), part I: 61-110,
        part II: 111-151.
\bibitem{be:quantal}
Berry,~M.V., Quantal phase factors accompanying adiabatic changes,
{\em Proc. R. Soc. London A} {\bf 392} (1984), 45-57.
\bibitem{co:dirac}
Cordes,~H.O., A pseudo-algebra of observables
 for the Dirac equation,
{\em Manuscripta Math.} {\bf 45} (1983), 77-105.
\bibitem{de:propagation}
Dencker,~N., On the propagation of polarization
sets for systems of real principal type, {\em
J. Funct. Anal.} {\bf 46} (1982), 351-372.
\bibitem {wil-lec:existence} {M. DeWilde, P.B.A. Lecomte:}
        {Existence of star-products and of formal deformations
        of the Poisson Lie Algebra of arbitrary 
         symplectic manifolds.}
        {\it Lett. Math. Phys.} {\bf 7} (1983), 487-496.
\bibitem{du:fourier}
  Duistermaat,~J. J., Fourier Integral Operators,{\em Courant
  Institute}, NYU, New York, 1973.
\bibitem{fedosov}  Fedosov,~B., 
         A Simple Geometrical Construction of 
               Deformation Quantization,
         J. of Diff. Geom. {\bf 40} (1994), 213-238.         
\bibitem{fedosov2}  Fedosov,~B.,
         Reduction and eigenstates in deformation quantization,
         In {\it Pseudo-Differential Calculus
                and Mathematical Physics,
          Advances in Prtial Differential Equations},
         Akademie Verlag, Berlin 1994, 227-297
\bibitem{gu-st:geometric}
  Guillemin,~V., and Sternberg,~S., Geometric Asymptotics, 
  {\em Math. Surveys} {\bf 14}, Amer. Math. Soc., Providence, 1977.
\bibitem{ho:fourier} 
H\"ormander,~L., Fourier Integral Operators I., {\em Acta Math.}
{\bf 127} (1971) 79-183.
\bibitem{ka:global}
Karasev,~M.V., 
       New global asymptotics and anomalies for the problem of
       quantization of the adiabatic invariant,
         {\em Funct. Anal. Appl.} {\bf 24}
       (1990), 104-114.
\bibitem{ka-ma:operators} 
Karasev,~M.V., and Maslov,~V.P., Operators with general commutation
relations and their applications.  I.  Unitary-nonlinear operator
equations, {\em J. Sov.  Math.} {\bf 15} (1981), 273-368.
\bibitem{klein}
Klein,~M, Martinez,~A, Wang~X.P.,
On the Born-Oppenheimer Approximation of Wave Operators in 
Molecular Scattering Theory,
{\em Comm. Math. Phys.} {\bf 152} (1993), 73-95
\bibitem{li-fl:geometric}
Littlejohn,~R.G., and Flynn,~W.G., 
Geometric phases in the asymptotic
theory of coupled wave equations,
{\em Phys. Rev.} {\bf A44} (1991), 5239-5256.
\bibitem{ma-fe:semiclassical}
Maslov,~V.P., and  Fedoriuk,~M.V.,
{Semi-classical Approximation in Quantum Mechanics},
D. Reidel, Dordrecht, 1981.
\bibitem{selbstwein} Emmrich,~C., and Weinstein,~A.,
Geometry of the transport equation in 
multicomponent WKB approximation,
 {\em Comm. Math. Phys.} {\bf 176} (1996), 701-711
\bibitem{yabana} Yabana,~K., Horiuchi,~H., 
Adiabatic Viewpoint for the WKB Treatment 
of Coupled Channel Systems,
{\em Prog. Theor. Phys.} {\bf 75} (1986), 592-618
\end{thebibliography}
 \end{document}